\documentclass[12pt]{article}
\usepackage{graphicx,amsmath,hyperref}
\def\hybrid{\topmargin 0pt      \oddsidemargin 0pt
        \headheight 0pt \headsep 0pt
        \voffset=-0.5cm
        \textwidth 6.25in       
        \textheight 9.5in       
        \marginparwidth 0.0in
        \parskip 5pt plus 1pt   \jot = 1.5ex}
\catcode`\@=11
\def\marginnote#1{}

\newcount\hour
\newcount\minute
\newtoks\amorpm
\hour=\time\divide\hour by60
\minute=\time{\multiply\hour by60 \global\advance\minute by-\hour}
\edef\standardtime{{\ifnum\hour<12 \global\amorpm={am}%
        \else\global\amorpm={pm}\advance\hour by-12 \fi
        \ifnum\hour=0 \hour=12 \fi
        \number\hour:\ifnum\minute<10 0\fi\number\minute\the\amorpm}}
\edef\militarytime{\number\hour:\ifnum\minute<10 0\fi\number\minute}

\def\draftlabel#1{{\@bsphack\if@filesw {\let\thepage\relax
   \xdef\@gtempa{\write\@auxout{\string
      \newlabel{#1}{{\@currentlabel}{\thepage}}}}}\@gtempa
   \if@nobreak \ifvmode\nobreak\fi\fi\fi\@esphack}
        \gdef\@eqnlabel{#1}}
\def\@eqnlabel{}
\def\@vacuum{}
\def\draftmarginnote#1{\marginpar{\raggedright\scriptsize\tt#1}}
\def\draftlabel#1{{\@bsphack\if@filesw {\let\thepage\relax
   \xdef\@gtempa{\write\@auxout{\string
      \newlabel{#1}{{\@currentlabel}{\thepage}}}}}\@gtempa
   \if@nobreak \ifvmode\nobreak\fi\fi\fi\@esphack}
        \gdef\@eqnlabel{#1}}
\def\@eqnlabel{}
\def\@vacuum{}
\def\draftmarginnote#1{\marginpar{\raggedright\scriptsize\tt#1}}

\def\draft{\oddsidemargin -.5truein
        \def\@oddfoot{\sl preliminary draft \hfil
        \rm\thepage\hfil\sl\today\quad\militarytime}
        \let\@evenfoot\@oddfoot \overfullrule 3pt
        \let\label=\draftlabel
        \let\marginnote=\draftmarginnote
   \def\@eqnnum{(\theequation)\rlap{\kern\marginparsep\tt\@eqnlabel}%
\global\let\@eqnlabel\@vacuum}  }

\def\numberbysection{\@addtoreset{equation}{section}
        \def\theequation{\thesection.\arabic{equation}}}

\def\underline#1{\relax\ifmmode\@@underline#1\else
        $\@@underline{\hbox{#1}}$\relax\fi}

\def\titlepage{\@restonecolfalse\if@twocolumn\@restonecoltrue\onecolumn
     \else \newpage \fi \thispagestyle{empty}\c@page\z@
        \def\thefootnote{\fnsymbol{footnote}} }

\def\endtitlepage{\if@restonecol\twocolumn \else  \fi
        \def\thefootnote{\arabic{footnote}}
        \setcounter{footnote}{0}}  
\relax

\hybrid

\newfont{\Bbb}{msbm10 scaled 1\@ptsize00}
\newfont{\Bbbb}{msbm7 scaled 1\@ptsize00}
\newcommand{\CC}{\mbox{\Bbb C}}

\newcommand{\DDD}{\raise-1pt\hbox{$\mbox{\Bbbb D}$}}

\newcommand{\UUU}{\raise-1pt\hbox{$\mbox{\Bbbb U}$}}

\newcommand{\z}{\raise-1pt\hbox{$\mbox{\Bbbb Z}$}}

\def\beq{\begin{equation}}
\def\eeq{\end{equation}}
\def\p{\partial}

\begin{document}

\begin{titlepage}

\title{Quantum spin chains and integrable 
many-body systems of classical mechanics}

\author{A. Zabrodin
\thanks{Institute of Biochemical Physics,
4 Kosygina, 119334, Moscow, Russia; ITEP, 25 B. Cheremushkinskaya,
117218, Moscow, Russia; National Research University Higher
School of Economics, International Laboratory of Representation 
Theory and Mathematical Physics,
20 Myasnitskaya Ulitsa, Moscow 101000, Russia}}

\date{September 2014}
\maketitle

\vspace{-7cm} \centerline{ \hfill ITEP-TH-29/14}\vspace{7cm}

\begin{abstract}

This note is a review of the recently revealed intriguing 
connection between integrable quantum spin chains 
and integrable many-body systems of classical mechanics.
The essence of this connection lies in the fact that the spectral 
problem for 
quantum Hamiltonians of the former models 
is closely related to a sort of inverse spectral problem for 
Lax matrices of the latter ones.
For simplicity, we focus on the most transparent and familiar case of
spin chains on $N$ sites constructed by means of the $GL(2)$-invariant
$R$-matrix. They are related to the classical Ruijsenaars-Schneider
system of $N$ particles, which is known to be an integrable deformation 
of the Calogero-Moser system. As an explicit example the case
$N=2$ is considered in detail.

\end{abstract}

\end{titlepage}

\vspace{5mm}

\vspace{5mm}


{\small
\section{Introduction}

In this paper we present some results of \cite{AKLTZ11}-\cite{GZZ14} in a 
short compressed form and in the simplest possible setting. 
First of all let us explain what we mean by ``quantum spin chains''
and ``integrable many-body systems of classical mechanics''.

The best known example of integrable quantum spin chain is the isotropic (XXX)
homogeneous Heisenberg model with spin $\frac{1}{2}$ on an 1D lattice
with coupling between nearest neighbours. 
Throughout the paper, we use the words ``spin chain'' in a broader sense,
not implying existence of any local Hamiltonian of the Heisenberg type.
In fact integrable local Hamiltonians in general do
not exist for inhomogeneous spin chains which 
are closely involved in our story.
However, such models still make sense as generalized spin chains with
long-range interaction and a family of commuting (non-local) 
Hamiltonians. We call them inhomogeneous XXX spin chains.
Alternatively, one may prefer to keep in mind 
inhomogeneous integrable lattice models of statistical
mechanics rather than spin chains as such.
In either case, the final goal of the theory is diagonalization of 
transfer matrices which are
generating functions of commuting conserved quantities.
This is usually achieved by one or another version of the 
Bethe ansatz method.

The integrable model of classical mechanics we are mainly interested 
in is the 
$N$-body system of particles on the line called the
Ruijsenaars-Schneider (RS) model \cite{RS}. 
It is often referred to as an integrable 
relativistic deformation of the famous Calogero-Moser (CM)
model with inversely quadratic pair potential \cite{Calogero,OP}. 

As is common for integrable models, the classical dynamics can be represented 
in the Lax form, i.e., as an isospectral deformation of 
a $N\! \times \! N$ matrix called the Lax matrix.
Matrix elements of this matrix
are simple functions of coordinates and momenta of the particles
while the eigenvalues are integrals of motion. 
In a nutshell, the essence of the quantum-classical (QC) duality
\beq\label{one}
\mbox{\fbox{Quantum integrable models} $\longleftrightarrow$
\fbox{Classical many-body systems}}\, .
\eeq
lies in the fact that
spectra of quantum Hamiltonians of a model from the left hand side
appear to be encoded 
in the algebraic properties of the Lax matrix for a classical system
from the right hand side. 

In the case of the inhomogeneous 
XXX spin chain, a refined version of (\ref{one}) is
\beq\label{two}
\mbox{\fbox{Quantum XXX spin-$\frac{1}{2}$ chain on $N$ sites} $\longleftrightarrow$
\fbox{Classical $N$-body RS model}}\, .
\eeq
More precisely,
the spectral problem for the quantum Hamiltonians of the 
inhomogeneous XXX spin chain on $N$ sites is reduced 
to a sort of an {\it inverse} spectral problem for the 
$N\! \times \! N$ Lax matrix for the classical RS system.
Given its spectrum and the coordinates of the particles, 
the problem is to find possible values of their momenta compatible
with these data. In general this problem has many solutions which 
just yield different eigenvalues of the quantum Hamiltonians.
In a special scaling limit, the XXX spin chain turns into the 
Gaudin spin model \cite{Gaudin}. On the right hand side of 
(\ref{two}), this corresponds to the non-relativistic limit
of the RS system:
\beq\label{three}
\mbox{\fbox{Quantum Gaudin model} $\longleftrightarrow$
\fbox{Classical CM model}}\,.
\eeq

The QC duality is traced back to \cite{GK}, where joint spectra
of some finite-dimensional operators were linked to the classical 
Toda chain. 
The existence of an unexpected link between the quantum Gaudin and the 
classical CM models 
was first pointed out in \cite{MTV09}, see also 
\cite{MTV12}.
In a more general set-up, the correspondence 
between quantum and classical integrable systems was 
independently derived \cite{AKLTZ11,Zsigma,Z12,Zjapan} 
as a corollary of 
an embedding of the commutative algebra of spin chain Hamiltonians 
into an infinite integrable hierarchy 
of soliton equations known as the modified Kadomtsev-Petviashvili (mKP)
hierarchy. Namely, the most general generating function of commuting integrals
of motion of the spin chain (the ``master $T$-operator'') was shown to satisfy
the bilinear identity and the Hirota bilinear 
equations for the tau-function of the mKP hierarchy \cite{DJKM83}. 

Although only a limited number of examples are available at the moment,
the very phenomenon of the existence 
of hidden non-standard connections between quantum
and classical integrable systems seems to be rather general.
Presumably, it can be thought of as a new kind of a correspondence (or duality)
principle in the realm of integrable systems. 
In \cite{GZZ14}, the QC duality 
(\ref{two}), (\ref{three}) was checked directly
using the Bethe ansatz solution of integrable spin chains. 
The role of this duality
in the context of supersymmetric gauge theories and branes 
was discussed in \cite{NRS,GK13,GZZ14}.

It is worthwhile to stress that the both sides of the correspondence, i.e.
quantum and classical integrable systems, participate 
in the game as two faces of one entity on an equal-rights basis. 
In the theory of quantum models, there are 
some fundamental relations, exact for any $\hbar \neq 0$,
which assume the form of classical equations of motion for some other system. 
(One of such examples is the classical integrable dynamics 
naturally realized in the 
space of conserved quantities of quantum integrable models, see
\cite{AKLTZ11} and earlier works \cite{KLWZ97,KSZ08}.)
At the same time, given a many-body problem of
classical mechanics, one may extract from it, by addressing some
non-traditional questions about the system, the spectral properties
of a quantum model.
This picture becomes valid and meaningful if the 
systems from both sides are integrable. 
It might be interesting to combine the hypothetical 
``correspondence principle'' based on the QC duality with 
the standard correspondence principle of quantum mechanics.

Let us outline the contents of the paper.

In section 2, we start with the most familiar
example of integrable spin chain: the 
Heisenberg model with spin 
$\frac{1}{2}$ and periodic boundary conditions (the XXX magnet)
solved by H.Bethe in 1931 \cite{Bethe}.
The ``spin variables'' are
vectors from the spaces $\CC^2$ at each site.
However, this model itself is too degenerate to be 
directly linked to a classical many-body system.
To this end, we need an inhomogeneous version of the model
with twisted boundary conditions. 
Such a generalized XXX model has $N+2$ free 
parameters which are $N$ ``inhomogeneity parameters'' on each site
and $2$ eigenvalues of the twist matrix which is assumed to be diagonal.
The generalized XXX model
can be naturally constructed in the 
framework of the Quantum Inverse Scattering Method (QISM)
developed by the former Leningrad school \cite{QISM1,FTLOMI}.
In the inhomogeneous 
model, the locality of spin interactions does not take place. 
Instead, there are
$N$ non-local commuting Hamiltonians (which are 
cousins of the Gaudin ones). They can be simultaneously 
diagonalized using the algebraic Bethe ansatz. 

In section 3, the necessary formulae related to the classical 
RS model are presented, including the Lax matrix.
The rules of the quantum-classical correspondence between the 
integrable models are explained in section 4.
As an example we consider the case $N=2$, where all calculations
can be done directly by hands (section 5). Finally, in section 6 
we give some remarks on the scaling limit to the Gaudin model
which corresponds, on the classical side, to the non-relativistic
limit of the RS system. Some generalizations and perspectives are
briefly discussed in the concluding section 7.

\section{The Heisenberg spin chain and its generalizations}

The Hamiltonian of the isotropic Heisenberg spin chain (also called the XXX-magnet)
with periodic boundary condition is
$$
{\bf H}^{\rm xxx}=2\sum_{j=1}^N \left 
({\bf s}_{x}^{(j)} {\bf s}_{x}^{(j+1)}+
{\bf s}_{y}^{(j)}{\bf s}_{y}^{(j+1)}+
{\bf s}_{z}^{(j)}{\bf s}_{z}^{(j+1)}-{\bf I}\right ),
\qquad N+1\equiv 1,
$$
where the spin operators $({\bf s}_{x}, {\bf s}_{y}, {\bf s}_{z})=
\vec {\bf s}$ are expressed through the Pauli matrices 
as
$$
{\bf s}_{x} =\frac{1}{2}\left (\begin{array}{cc} 0 &1 \\ 1& 0 \end{array}
\right ), \quad 
{\bf s}_{y} =\frac{1}{2}\left (\begin{array}{cc} 0 &-i \\ i& 0 \end{array}
\right )\quad 
{\bf s}_{z}=\frac{1}{2}\left (\begin{array}{cc} 1 &0 \\ 0& -1 \end{array}
\right )
$$
and ${\bf I}={\bf 1}^{\otimes N}$ is the identity operator.
(Hereafter ${\bf 1}$ stands for the identity matrix in $\CC^2$).
We will also use ${\bf s}_+={\bf s}_x +i {\bf s}_y=
\left ( \begin{array}{cc}
0&1\\0&0\end{array}\right )$, $ {\bf s}_-={\bf s}_x -i {\bf s}_y=
\left ( \begin{array}{cc}
0&0\\1&0\end{array}\right )$, 
${\bf s}_1=\frac{1}{2} {\bf 1} +{\bf s}_z=
\left ( \begin{array}{cc}
1&0\\0&0\end{array}\right )$ and
${\bf s}_2=\frac{1}{2} {\bf 1} -{\bf s}_z=
\left ( \begin{array}{cc}
0&0\\0&1\end{array}\right )$.
The operator
$\vec {\bf s}^{(j)}= {\bf 1}^{\otimes (j-1)}\otimes \vec {\bf s}\otimes 
{\bf 1}^{\otimes (N-j)}$ acts non-trivially at the $j$th site 
of the chain.
Clearly, they commute for  any $j'\ne j$. 
The Hamiltonian acts in the $2^N$-dimensional linear space 
${\cal V}=\otimes_{j=1}^N V_j$, $V_j \cong \CC^2$.
Basis vectors
in this space can be constructed as tensor products of local vectors with
definite $z$-projection of spin, i.e., eigenvectors of ${\bf s}_z$. 

Note that
$
{\bf P}_{ij}=\frac{1}{2}\bigl ({\bf I}+4\vec {\bf s} ^{(i)}\vec {\bf s} ^{(j)}\bigr )
$
is the permutation operator of the $i$th and $j$th spaces, and so 
the Heisenberg Hamiltonian can be written in the form
${\bf H}^{\rm xxx}=\sum_j {\bf P}_{j\,\, j+1} \, - \, N{\bf I}$. 

The Hamiltonian commutes with the operator
\beq\label{I1}
{\bf M}=\frac{1}{2}\sum_{j=1}^N ({\bf I}-2{\bf s}_z ^{(j)})=
\sum_{j=1}^N {\bf s}_2 ^{(j)}
\eeq
which counts the total number of spins in the chain with negative
$z$-projection. Namely,
the states in which $M$ spins look down (and so the rest $N-M$ spins look up)
are eigenstates for the operator ${\bf M}$ with the eigenvalue $M$.
The space of states ${\cal V}$ is decomposed in the direct sum of 
eigenspaces for the operator ${\bf M}$: 
$\displaystyle{{\cal V}=\bigoplus_{M=0}^{N}{\cal V}(M)}$,
${\bf M}{\cal V}(M)=M{\cal V}(M)$. It is clear that
$$
\mbox{dim}\, {\cal V}(M) =\left (\begin{array}{c}
N \\ M\end{array}\right ) =\frac{N!}{M! (N\! -\! M)!}\,.
$$
In particular, ${\cal V}(0)$ and ${\cal V}(N)$ are one-dimensional 
spaces generated by the states in which all spins look up or down
respectively.

The common spectral problem for the operators ${\bf H}^{\rm xxx}$ and ${\bf M}$,
${\bf H}^{\rm xxx}\Psi =E\Psi$,  ${\bf M}\Psi =M\Psi$, has the famous 
Bethe ansatz solution \cite{Bethe}. The eigenvalues $E$
for $0\leq M\leq [N/2]$ are given by the formula
\beq\label{E}
E=\sum_{\alpha =1}^{M} \varepsilon (v_{\alpha}), \quad
\varepsilon (v)=-\, \frac{4}{1+4v^2}\,,
\eeq
where the auxiliary quantities $v_{\alpha}$ (the Bethe roots) are to be
found from the system of algebraic equations
\beq\label{Bethe1}
\left (\frac{v_\alpha +\frac{i}{2}}{v_\alpha -\frac{i}{2}}\right )^N
=\prod_{\beta =1, \beta \neq \alpha}^{M}
\frac{v_\alpha -v_\beta +i}{v_\alpha -v_\beta -i}
\eeq
(the Bethe equations). Different solutions to this system give energies
of different eigenstates.

The exact solution of the Heisenberg spin chain is possible
due to the fact that the model is integrable. This means that 
there is a sufficiently large family of independent commuting operators,
one of which is the Heisenberg Hamiltonian. The other operators of 
this family are higher integrals of motion. A general prescription how to
construct models possessing higher integrals of motion is provided by the 
Quantum Inverse Scattering Method (QISM) \cite{QISM1}.

We start by reformulating the XXX spin chain in the 
framework of the QISM, following \cite{FTLOMI}. Such a reformulation 
makes integrability of the model explicit and, what is even more 
important, it suggests natural integrable generalizations of the XXX chain.

Let $V_0\cong \CC^2$ be another copy of the complex linear space $\CC^2$
(the auxiliary space). 
The quantum Lax operator at the $j$th site acts non-trivially in
$V_0\otimes V_j$. It is
\beq\label{J00}
{\sf L}_{j}(x)=x {\bf 1}\otimes {\bf I}+\eta {\bf P}_{0 j}=
\Bigl ( x+\frac{\eta}{2}\Bigr ) {\bf 1}\otimes {\bf I}
+2\eta \, \vec {\bf s} \otimes \vec {\bf s}\,,
\eeq
or, in the block-matrix form,
\beq\label{J0}
{\sf L}_{j}(x)=\left ( \begin{array}{cc}
x{\bf I}+ \eta {\bf s}_1^{(j)} & \eta \, {\bf s}_-^{(j)} \\ & \\
\eta \, {\bf s}_+^{(j)} & x{\bf I}+\eta {\bf s}_2^{(j)}
\end{array} \right ).
\eeq
The variable $x\in \CC $ is called the (quantum) spectral parameter.
The extra parameter $\eta$ introduced here 
for the reason clarified below
is not actually essential because it can be
eliminated by a rescaling of the spectral parameter 
(unless one tends it to 0 as in the limit to the Gaudin model 
\cite{Gaudin}). The Heisenberg 
Hamiltonian does not depend on $\eta$ which is usually put equal to 
$i=\sqrt{-1}$
in this context.
The $L$-operator satisfies the ``$RLL=LLR$'' intertwining relation 
$$
{\sf R}(x-x')\, {\sf L}_{j}(x)\otimes {\sf L}_{j}(x') ={\sf L}_{j}(x')
\otimes {\sf L}_{j}(x)\, {\sf R}(x-x'),
$$
where the quantum $R$-matrix ${\sf R}(x)$ acts in the tensor product of two 
auxiliary spaces $V_0 \cong V_{0'}\cong \CC^2$. In the natural basis 
in $\CC^2 \otimes \CC^2$ it is
\beq\label{J2}
{\sf R}(x)=\left (\begin{array}{cccc}
x\! +\! \eta & 0& 0 &0
\\ 
0& \eta & x & 0 
\\
0 & x & \eta & 0
\\
0& 0 &0 & x\! +\! \eta \end{array} \right ) =
\eta \, {\bf 1}\, \otimes {\bf 1}+x{\bf P}_{00'}.
\eeq
Note that in this particular case the $R$-matrix 
is almost 
the same object as the quantum $L$-operator: they differ 
only by a permutation operator of the two spaces, so that the 
intertwining relation 
is equivalent to the Yang-Baxter equation for the $R$-matrix.
The quantum transfer matrix is defined as
\beq\label{J1}
{\bf T}(x)=\mbox{tr}_0\Bigl [ {\sf L}_{1}(x)
{\sf L}_{2}(x)\, \ldots \, {\sf L}_{N}(x)\Bigr ]
= 2{\bf I} \, x^N  + {\bf J}_{N\! -\! 1}x^{N\! -\! 1}+\ldots + {\bf J}_1x +
{\bf J}_0\,.
\eeq
The intertwining relation implies that the transfer matrices 
with different spectral parameters (and the same $\eta$) 
commute: $[{\bf T}(x), \, {\bf T}(x')]=0$
for any $x, x'$. In its turn, this implies that the operators 
${\bf J}_k$ in (\ref{J1}) all commute with each other.
At the same time, the operator ${\bf J}_0$ is proportional to the 
cyclic permutation of the chain:
$$
{\bf J}_0={\bf T}(0)=\eta^N {\bf P}_{12} {\bf P}_{23} {\bf P}_{34}
\ldots {\bf P}_{N-1\,\, N}{\bf P}_{N1}
$$
while the Hamiltonian of the spin chain is given by
$$
{\bf H}^{\rm xxx} =\eta \frac{d}{dx}\, \log {\bf T}(x)\Bigr |_{x=0} -N{\bf I}=
\eta {\bf J}_0^{-1}{\bf J}_1 -N{\bf I}.
$$
The operators ${\bf J}_0^{-1}{\bf J}_k$ are then 
the higher integrals of motion.
The operator ${\bf J}_{0}^{-1}{\bf J}_1$ is local due to the special property of the quantum Lax operator
${\sf L}_{j}(0)=\eta {\bf P}_{0j}$ and the homogeneity of the chain.
The operator ${\bf M}$ (see (\ref{I1})) commutes not only with ${\bf H}^{\rm xxx}$
but with the whole one-parametric family ${\bf T}(x)$, and the Bethe states
are common eigenstates for the ${\bf T}(x)$ and ${\bf M}$:
${\bf T}(x)\Psi =T(x)\Psi$, ${\bf M}\Psi =M\Psi$.

The transfer matrix ${\bf T}(x)$ can be diagonalized by means of the 
algebraic Bethe ansatz method.
The eigenvalues $T(x)$ are given by the formula
\beq\label{J5}
T(x)=(x+\eta )^N \prod_{\alpha =1}^{M}\frac{x-u_\alpha -\eta}{x-u_\alpha}+
x^N \prod_{\alpha =1}^{M}\frac{x-u_\alpha +\eta}{x-u_\alpha}\,.
\eeq
The Bethe roots $u_{\alpha}$ are to be found from the 
system of Bethe equations
\beq\label{Bethe2}
\left (\frac{u_\alpha +\eta}{u_\alpha}\right )^N
=\prod_{\beta =1, \beta \neq \alpha}^{M}
\frac{u_\alpha -u_\beta +\eta}{u_\alpha -u_\beta -\eta}\,,
\eeq
where is implied that $0\leq M \leq [N/2]$.
The eigenvalues of the Heisenberg Hamiltonian, in terms of the Bethe roots,
are given by the formula
$$
E=\sum_{\alpha =1}^{M}\frac{\eta^2}{u_{\alpha}(u_{\alpha}+\eta )}
$$
which is equivalent to (\ref{E}) under the substitution
$\displaystyle{v_{\alpha}=\frac{iu_\alpha}{\eta}+\frac{i}{2}}$.

The XXX model can be generalized, preserving integrability, 
in two ways: a) by making it 
inhomogeneous and b) by imposing twisted boundary conditions.
The former is based on the possibility to introduce an inhomogeneity 
parameter at each site which does not spoil the intertwining relation:
$$
{\sf R}(x-x')\, {\sf L}_{j}(x-x_j)\otimes {\sf L}_{j}(x'-x_j) 
={\sf L}_{j}(x'-x_j)
\otimes {\sf L}_{j}(x-x_j)\, {\sf R}(x-x').
$$
The latter is due to the $GL(2)$-invariance of the $R$-matrix
(\ref{J2}):
$
g \otimes g \, {\sf R}(x)={\sf R}(x) \, g \otimes g
$
for any $g\in GL(2)$. This property implies that
commutativity of the transfer matrices still holds if one inserts
a matrix $g\in GL(2)$ in the auxiliary space before taking trace.
For simplicity, we assume that $g$ is diagonal:
\beq\label{g}
g=\left ( \begin{array}{cc} w_1 & 0\\ 0& w_2\end{array} \right ).
\eeq
The generalizations a) and b) 
can be applied simultaneously, which leads 
to the most general one-parametric family 
of commuting operator-valued polynomials in $x$:
\beq\label{J3}
{\bf T}(x)={\bf T}(x;g,\eta , \{x_j\})= 
\mbox{tr}_0\Bigl [ g \, {\sf L}_{1}(x-x_1)
{\sf L}_{2}(x-x_2)\, \ldots \, {\sf L}_{N}(x-x_N)\Bigr ].
\eeq
These operators commute for different $x$'s and the same 
$\eta$, $g$ and $x_j$: $$[{\bf T}(x;g,\eta , \{x_j\}),\, 
{\bf T}(x';g,\eta , \{x_j\})]=0.$$
Similarly to (\ref{J1}), one can expand
\beq\label{J4}
{\bf T}(x)
= {\bf I} \, \mbox{tr}\, g \,
x^N  + {\bf J}_{N\! -\! 1}x^{N\! -\! 1}+\ldots + {\bf J}_1x +
{\bf J}_0,
\eeq
the ${\bf J}_k$'s being commuting integrals of motion.
Note, in particular, that 
${\bf J}_{N-1}=\eta \sum_{i}{\bf g}^{(i)}$, where
${\bf g}^{(i)}$ is the operator acting as the matrix $g$ at the $i$th site:
${\bf g}^{(i)}:= {\bf 1}^{\otimes (i-1)} \otimes g \otimes 
{\bf 1}^{\otimes (N-i)}$.
In general 
there is no way to construct local Hamiltonians from the
${\bf J}_k$'s.
Instead, assuming that 
all the $x_j$'s are distinct and in general position
(meaning that $x_i-x_j \neq \pm \eta$ for all $i,j$), 
one can define {\it non-local} Hamiltonians as residues 
of ${\bf T}(x)/\prod_j (x-x_j)$ (cf. \cite{HKW92}):
$$
\frac{{\bf T}(x)}{\prod_{j=1}^N (x-x_j)}=
\mbox{tr}\, g \cdot {\bf I}\,  +\sum_{j=1}^{N}\frac{\eta \, {\bf H}_j}{x-x_j}\,.
$$
In general, the Hamiltonians 
${\bf H}_j={\bf H}_j(\eta , g, \{x_i\})$ 
imply a long-range interaction involving all spins in
the chain. Their explicit form is
\beq\label{Hi}
{\bf H}_i=\overrightarrow{\prod_{j=i+1}^{N}}\left (
{\bf I}+\frac{\eta \, {\bf P}_{ij}}{x_i-x_j}\right ){\bf g}^{(i)}\,
\overrightarrow{\prod_{\,\, j=1\,\,}^{i-1}}\left (
{\bf I}+\frac{\eta \,{\bf P}_{ij}}{x_i-x_j}\right ),
\eeq
where 
we use the notation 
$\displaystyle{\overrightarrow{\prod_{j=1}^{m}}A_j =
A_1A_2\ldots A_m}$ for the ordered product. 
It follows from the definition that
$\displaystyle{\sum_{j=1}^N {\bf H}_j=\sum_{j=1}^N {\bf g}^{(j)}}$.

The operator ${\bf M}$ (\ref{I1}) still commutes with
${\bf T}(x)$ and all the ${\bf H}_j$'s, so, again, 
all these operators are diagonalized simultaneously:
${\bf T}(x)\Psi =T(x)\Psi$, 
${\bf H}_j \Psi = H_j \Psi$,
${\bf M}\Psi =M\Psi$.
The algebraic Bethe ansatz gives the following result.
The eigenvalues $T(x)$ and $H_j$ are given by the formulae
\beq\label{J5a}
T(x)=w_1\prod_{k=1}^N (x-x_k+\eta )  \prod_{\alpha =1}^{M}
\frac{x-u_\alpha -\eta}{x-u_\alpha}+
w_2\prod_{k=1}^N (x-x_k) 
\prod_{\alpha =1}^{M}\frac{x-u_\alpha +\eta}{x-u_\alpha},
\eeq
\beq\label{J6}
H_j= w_1 \prod_{k=1, \neq j}^N \frac{x_j\! -\! x_k \! 
+\! \eta}{x_j \! -\! x_k}\, \prod_{\alpha =1}^{M}
\frac{x_j\! -\! u_\alpha \! 
-\! \eta}{x_j \! -\! u_\alpha }.
\eeq
The Bethe roots $u_{\alpha}$ are to be found from the 
system of Bethe equations
\beq\label{Bethe3}
\frac{w_1}{w_2}\prod_{k=1}^N \frac{u_\alpha -x_k +\eta}{u_\alpha -x_k}
\, =\prod_{\beta =1, \beta \neq \alpha}^{M}
\frac{u_\alpha -u_\beta +\eta}{u_\alpha -u_\beta -\eta}\,, 
\eeq
where it is implied that $0\leq M \leq [N/2]$.

\section{The Ruijsenaars-Schneider model}

The RS model \cite{RS} is an integrable model of classical mechanics. 
It is an $N$-body system of interacting
particles on the line with the Hamiltonian
\beq\label{RS1}
{\cal H}_1^{\rm RS}=\eta ^{-1}\sum_{i=1}^N e^{-\eta p_i}\prod_{k=1, \neq i}^N 
\frac{x_i-x_k+\eta}{x_i-x_k}\,.
\eeq
For some reason it is often called the relativistic deformation 
of the Calogero-Moser model, the parameter $\eta$ being the inverse 
``velocity of light''. The Hamiltonian equations of motion 
$\displaystyle{\left ( \begin{array}{c}\dot x_i \\ \dot p_i \end{array} \right )
=\left ( \begin{array}{cc} \p_{p_i}{\cal H}_1^{\rm RS}\\
-\p_{x_i}{\cal H}_1^{\rm RS}\end{array} \right )}$ give the following
connection between velocity and momentum
\beq\label{RS2a}
\dot x_i=-e^{-\eta p_i}\prod_{k=1, \neq i}^N 
\frac{x_i-x_k+\eta}{x_i-x_k}
\eeq
and the equations of motion
\beq\label{RS2}
\ddot x_i = -\sum_{k\neq i}\frac{2\, \eta^2 \dot x_i \dot x_k}{(x_i-x_k)
((x_i-x_k)^2 -\eta^2)}\,, \qquad i=1, \ldots , N.
\eeq

The RS model is known to be integrable, with the higher integrals
of motion in involution being given by ${\cal H}^{\rm RS}_k=
\eta^{-1}\mbox{tr}\, ({\sf Y}^{\rm RS})^k$, where ${\sf Y}^{\rm RS}=
{\sf Y}^{\rm RS}(\{x_i\}; \{\dot x_i\})$ is the Lax matrix of the model.
Its matrix elements are $\displaystyle{{\sf Y}^{\rm RS}_{ij} 
=\frac{\eta \dot x_i}{x_i-x_j-\eta}}$, i.e., 
\beq\label{Lax1}
{\sf Y}^{\rm RS}(\{x_i\}; \{\dot x_i\})
=\left ( \begin{array}{ccccc}
\displaystyle{-\dot x_1} & \displaystyle{\frac{\eta \dot x_1}{x_1\! -\!x_2\!-\! \eta}} 
&\displaystyle{\frac{\eta \dot x_1}{x_1\! -\! x_3\! -\! \eta}} &
\ldots & \displaystyle{\frac{\eta \dot x_1}{x_1\! -\! x_N\! -\! \eta}}
\\ &&&& \\
 \displaystyle{\frac{\eta \dot x_2}{x_2\! -\! x_1\! -\! \eta}} & 
 \displaystyle{-\dot x_2}& 
 \displaystyle{\frac{\eta \dot x_2}{x_2\! -\! x_3\! -\! \eta}} &
 \ldots & \displaystyle{\frac{\eta \dot x_2}{x_2\! -\! x_N\! -\! \eta}}
 \\ &&&& \\ \vdots & \vdots & \vdots & \ddots & \vdots
 \\ &&&& \\
 \displaystyle{\frac{\eta\dot x_N}{x_N\! -\! x_1\! -\! \eta}} & 
 \displaystyle{\frac{\eta\dot x_N}{x_N\! -\! x_2\! -\! \eta}}&
 \displaystyle{\frac{\eta\dot x_N}{x_N\! -\! x_3\! -\! \eta}} & \ldots & 
 \displaystyle{-\dot x_N}
\end{array}\right ).
\eeq
Equations of motion (\ref{RS2}) are equivalent to the Lax equation
$
\dot {\sf Y}^{\rm RS}=[{\sf B}, {\sf Y}^{\rm RS}]
$,
where
$$
{\sf B}_{ij}=\left (\sum_{k\neq i}\frac{\dot x_k}{x_i-x_k}-
\sum_k \frac{\dot x_k}{x_i-x_k+\eta }\right )\delta_{ij}
+\frac{\dot x_i}{x_i-x_j}\, (1-\delta_{ij}).
$$
The Lax equation implies that all eigenvalues of the Lax matrix
are integrals of motion.

Let ${\sf X}=\mbox{diag}(x_1, x_2, \ldots , x_N)$ be the diagonal matrix
with the diagonal entries being coordinates of the particles. It is easy to check that
the matrices ${\sf X}$, ${\sf Y}^{\rm RS}$ satisfy the commutation relation
\beq\label{RS3}
[{\sf X},{\sf Y}^{\rm RS}]=\eta {\sf Y}^{\rm RS}+\eta \dot {\sf X}{\sf E},
\eeq
where ${\sf E}$ is the $N\! \times \! N$ matrix of 
rank $1$ with all entries equal to $1$.
Note also that the Lax matrix ${\sf Y}^{\rm RS}$ can be represented 
in the form
\beq\label{RS4}
{\sf Y}^{\rm RS}=\dot {\sf X}\, {\sf C},
\eeq
where ${\sf C}$ is the Cauchy matrix 
$\displaystyle{{\sf C}_{ij}=\frac{\eta}{x_i-x_j-\eta}}$.

\section{The quantum-classical duality}

Consider the Lax matrix (\ref{Lax1}) of the 
$N$-particle RS model, where the $x_i$'s
are identified 
with the inhomogeneity parameters $x_i$ at the sites of the spin chain
and the inverse ``velocity of light'', $\eta$, is identified with the 
parameter $\eta$ introduced in the quantum $L$-operator (\ref{J0}).
Let us also substitute 
$\dot x_i =-H_i$:
\beq\label{QC1}
{\sf Y}^{\rm RS}(\{x_i\}; \{-H_i\})
=\left ( \begin{array}{ccccc}
\displaystyle{H_1} & \displaystyle{\frac{\eta H_1}{x_2\! -\!x_1\!+\! \eta}} 
&\displaystyle{\frac{\eta H_1}{x_3\! -\! x_1\! +\! \eta}} &
\ldots & \displaystyle{\frac{\eta H_1}{x_N\! -\! x_1\! +\! \eta}}
\\ &&&& \\
 \displaystyle{\frac{\eta H_2}{x_1\! -\! x_2\! +\! \eta}} & 
 \displaystyle{H_2}& 
 \displaystyle{\frac{\eta H_2}{x_3\! -\! x_2\! +\! \eta}} &
 \ldots & \displaystyle{\frac{\eta H_2}{x_N\! -\! x_2\! +\! \eta}}
 \\ &&&& \\ \vdots & \vdots & \vdots & \ddots & \vdots
 \\ &&&& \\
 \displaystyle{\frac{\eta H_N}{x_1\! -\! x_N\! +\! \eta}} & 
 \displaystyle{\frac{\eta H_N}{x_2\! -\! x_N\! +\! \eta}}&
 \displaystyle{\frac{\eta H_N}{x_3\! -\! x_N\! +\! \eta}} & \ldots & 
 \displaystyle{H_N}
\end{array}\right ).
\eeq
The decomposition (\ref{RS4}) for the matrix (\ref{QC1})
acquires the form
\beq\label{RS4a}
{\sf Y}^{\rm RS}(\{x_i\}; \{-H_i\})=-{\sf H}{\sf C},
\eeq
where ${\sf H}=\mbox{diag}(H_1, H_2, \ldots , H_N)$.

The claim is that if the $H_i$'s are eigenvalues of the Hamiltonians
of the spin chain in the invariant subspace ${\cal V}(M)$, then 
the first $N-M$ eigenvalues of this matrix coincide with
eigenvalues of the twist matrix $w_1$ 
while the rest $M$ eigenvalues 
coincide with $w_2$:
\beq\label{QC2}
\mbox{Spec}\, ({\sf Y}^{\rm RS})=\Bigl (\underbrace{w_1, \ldots , w_1}_{N-M}, \,
\underbrace{w_2, \ldots , w_2}_{M} \Bigr ).
\eeq
This means that the values of the higher RS Hamiltonians are
\beq\label{QC3}
\eta {\cal H}^{\rm RS}_k=(N-M)w_1^k+Mw_2^k.
\eeq
In general, the matrix ${\sf Y}^{\rm RS}$ with multiple eigenvalues 
is not diagonalizable and contains Jordan cells.

To put it somewhat differently, one can say that the eigenstates 
of the quantum spin chain Hamiltonians
correspond to the intersection points of two Lagrangian submanifolds
in the phase space of the RS model. One of them is the hyperplane
defined by fixing all the coordinates $x_i$ while the other one is
the Lagrangian submanifold obtained by fixing values (\ref{QC3})
of the $N$ integrals of motion in involution ${\cal H}^{\rm RS}_k$.
In general, there are many such intersection points numbered 
by a finite set $I$, with coordinates,
say $(x_1, \ldots , x_N, \, 
p_1^{(\alpha )}, \ldots , p_N^{(\alpha )})$, $\alpha \in I$.
The values of $p_j^{(\alpha )}$ give, through equation (\ref{RS2a}),
the spectrum of ${\bf H}_j$:
$$
H_j^{(\alpha )} =e^{-\eta p_j^{(\alpha )}}\prod_{k=1, \neq j}
\frac{x_j-x_k +\eta}{x_j-x_k}\,.
$$
However, we can not claim that all the intersection points 
correspond to the energy levels of the spin chain Hamiltonians. 
The example of $N=2$ considered below in detail suggests that
some intersection points do not correspond to the energy levels 
of a given spin chain. Their meaning is to be clarified.

Anyway,
the spectral problem for the non-local inhomogeneous spin chain
Hamiltonians ${\bf H}_j$ in the subspace ${\cal V}(M)$ appears 
to be closely linked to the following {\it inverse spectral problem} 
for the RS Lax matrix ${\sf Y}^{\rm RS}$ 
of the form (\ref{QC1}).
Let us fix the spectrum of the matrix ${\sf Y}^{\rm RS}$ to be
(\ref{QC2}),
where $w_1$, $w_2$ are eigenvalues of the (diagonal) twist matrix $g$.
Then we ask what is the set of possible 
values of the $H_j$'s allowed by these constraints.
The eigenvalues $H_j$ of the quantum
Hamiltonians are contained in this set.

A similar correspondence between quantum and classical integrable
systems was suggested in \cite{MTV09}, see also 
\cite{MTV12}.
In a more general set-up, this assertion was derived \cite{AKLTZ11,Zsigma,Z12,Zjapan} 
as a corollary of 
the embedding of the spin chain into an infinite integrable hierarchy 
of non-linear PDE's. In \cite{GZZ14}, it was checked directly
using the Bethe ansatz solution.

In order to find the characteristic polynomial of the matrix
(\ref{QC1}) explicitly, 
we use the well known fact that
the coefficient in front of $\lambda^{N-k}$
in the polynomial $\det_{N\times N} (\lambda {\sf I}+{\sf A})$ equals
the sum of all diagonal $k\! \times \! k$ minors of the matrix ${\sf A}$.
All such minors can be found using
decomposition (\ref{RS4a}) and the explicit expression for the 
determinant of the Cauchy matrix:
$$
\det_{1\leq i,j \leq n}\, \frac{\eta}{x_i\! -\! x_j\! -\! \eta}=
(-1)^n \prod_{1\leq i<j \leq n}
\left (1-\frac{\eta^2}{(x_i\! -\! x_j)^2}\right )^{-1}.
$$
The result is:
\beq\label{QC4}
\det_{N\times N}(\lambda {\sf I}-{\sf Y}^{\rm RS})=
\det_{N\times N}(\lambda {\sf I}+{\sf H}{\sf C})=
\sum_{n=0}^N {\cal J}_n \lambda ^{N-n},
\eeq
where
\beq\label{QC5}
{\cal J}_n = (-1)^n \!\!
\sum_{1\leq i_1<\ldots < i_n \leq N}
H_{i_1}\ldots H_{i_n}
\prod_{1\leq \alpha < \beta \leq n}
\left (1-\frac{\eta^2}{(x_{i_{\alpha}}\! -\! x_{i_{\beta}})^2}\right )^{-1}.
\eeq
In particular, the highest coefficient is given by the following
simple formula:
$$
{\cal J}_N=(-1)^N \, H_1 H_2 \ldots H_N \prod_{i<j}
\left (1-\frac{\eta^2}{(x_{i}\! -\! x_{j})^2}\right )^{-1}.
$$
For completeness, we point out that the integrals ${\cal H}_k$ introduced
in the previous section are connected with the integrals ${\cal J}_k$
by the Newton's formula \cite{Macdonald}
$\displaystyle{\sum_{k=0}^N {\cal J}_{N-k}{\cal H}_k=0}$ (we have set
${\cal H}_0=\eta^{-1}\mbox{tr} ({\sf Y}^{\rm RS})^0=N/\eta$).

Another way to write expressions (\ref{QC4}), (\ref{QC5}) is through a sum over
$\epsilon_1, \ldots , \epsilon_N$, with $\epsilon_i \in \{0,1\}$:
\beq\label{QC6}
\det_{N\times N}(\lambda {\sf I}-{\sf Y}^{\rm RS})=\lambda^N
\sum_{\{\epsilon_1, \ldots , \epsilon_N\}\in \z_2^N}
\prod_{i=1}^N \Bigl (-H_i/\lambda \Bigr )^{\epsilon_i}\prod_{1\leq j<k\leq N}
\left (1-\frac{\eta^2}{(x_j-x_k)^2}\right )^{-\epsilon_j \epsilon_k}.
\eeq
The similarity of these expressions with tau-functions
for $N$-soliton solutions to the KP hierarchy is not accidental.  
This point will be discussed elsewhere. 

We conclude this section by writing down the system of algebraic 
equations for spectra of the operators ${\bf H}_i$. 
Combining (\ref{QC2}) and (\ref{QC5}), we obtain $N$ polynomial equations 
for $N$ unknown quantities $H_1, \ldots , H_N$:
\beq\label{QC7}
\sum_{1\leq i_1<\ldots < i_n \leq N}
H_{i_1}\ldots H_{i_n}
\prod_{1\leq \alpha < \beta \leq n}
\left (1-\frac{\eta^2}{(x_{i_{\alpha}}\! -\! x_{i_{\beta}})^2}\right )^{-1}
=C_n(N,M),
\eeq
where $\displaystyle{C_n(N,M)=\frac{1}{2\pi i}\oint_{|z|=1}(1+zw_1)^{N-M}
(1+zw_2)^M z^{-n-1}dz}$, $n=1,2, \ldots , N$.
Let us emphasize that in contrast to the Bethe ansatz solution, 
the algebraic equations are written here 
not for some auxiliary quantities like Bethe roots but
for the spectrum itself. 

The state where all spins look up ($M=0$) is an obvious 
eigenvector of the operators ${\bf H}_i$ with the eigenvalues
\beq\label{QC8}
H_i=w_1\prod_{j=1, \neq i}^N \Bigl (1+\frac{\eta}{x_i-x_j}\Bigr ).
\eeq
One can check that these $H_i$'s indeed solve the system (\ref{QC7})
with $\displaystyle{C_n(N,0)=\frac{N! w_1^n}{n! (N-n)!}}$.

\section{Examples: $N=1$ and $N=2$}

The case $N=1$
is trivial. The only quantum Hamiltonian ${\bf H}_1$ is diagonal
in the standard basis of $\CC^2$ and coincides with the twist matrix,
so we have two eigenvalues: $H_1 =w_1$ or $H_1 =w_2$.
The one-particle RS model is the model of a free particle on the line,
the Lax ``matrix'' is just the number $-\dot x_1$. Fixing it to be
$w_1$ or $w_2$, as required by the QC duality, we obtain the 
two eigenvalues of ${\bf H}_1$ by the identification 
$H_i =-\dot x_i$, see (\ref{QC1}).

The case $N=2$ is meaningful and instructive.
First, let us find the spectrum of the quantum Hamiltonians directly.
The transfer matrix is:
$$
{\bf T}(x)\! =\! \mbox{tr} \!\! \left [
\left ( \!\!\begin{array}{cc}
w_1 & 0\\0& w_2 \end{array} \!\!\right )\!\!
\left (\!\! \begin{array}{cc} (x\! -\! x_1){\bf I} 
\! +\! \eta {\bf s}_{1}^{(1)}
& \eta \, {\bf s}_{-}^{(1)}\\
\eta \, {\bf s}_{+}^{(1)} & 
(x\! -\! x_1){\bf I} \! +\! \eta {\bf s}_{2}^{(1)}
\end{array} \!\!\!\!
\right ) \!\!
\left ( \!\!\!\! \begin{array}{cc} (x\! -\! x_2){\bf I} \! +\! \eta {\bf s}_{1}^{(2)}
& \eta \, {\bf s}_{-}^{(2)}\\
\eta \, {\bf s}_{+}^{(2)} & 
(x\! -\! x_2){\bf I} \! +\! \eta {\bf s}_{2}^{(2)}
\end{array} \!\!\!\!
\right )
\right ]
$$
A simple calculation gives the following explicit form of the
Hamiltonians:
$$
{\bf H}_1=w_1 {\bf s}_{1}^{(1)}+
w_2 {\bf s}_{2}^{(1)}+
\frac{\eta w_1}{x_1-x_2}
({\bf s}_{1}^{(1)}{\bf s}_{1}^{(2)}+
{\bf s}_{-}^{(1)}{\bf s}_{+}^{(2)})+
\frac{\eta w_2}{x_1-x_2}
({\bf s}_{2}^{(1)}{\bf s}_{2}^{(2)}+
{\bf s}_{+}^{(1)}{\bf s}_{-}^{(2)}),
$$
$$
{\bf H}_2=w_1 {\bf s}_{1}^{(2)}+
w_2 {\bf s}_{2}^{(2)}+
\frac{\eta w_1}{x_2-x_1}
({\bf s}_{1}^{(1)}{\bf s}_{1}^{(2)}+
{\bf s}_{-}^{(1)}{\bf s}_{+}^{(2)})+
\frac{\eta w_2}{x_2-x_1}
({\bf s}_{2}^{(1)}{\bf s}_{2}^{(2)}+
{\bf s}_{+}^{(1)}{\bf s}_{-}^{(2)}).
$$
We see that ${\bf H}_1 +{\bf H}_2 ={\bf g}^{(1)}+{\bf g}^{(2)}$,
as it should be. The space $\CC^2 \otimes \CC^2$ is decomposed into
the direct sum of the one-dimensional space ${\cal V}(0)$ generated by the
vector $\left | ++\right >$ ($M=0$), two-dimensional space ${\cal V}(1)$
generated by the
vectors $\left | +-\right >$, $\left | -+\right >$ ($M=1$) and
one-dimensional space ${\cal V}(2)$ generated by the
vector $\left | --\right >$ ($M=2$).
We have:
$$
{\bf H}_1 \left | ++\right >=w_1\Bigl (1+\frac{\eta}{x_1-x_2}\Bigr )
\left | ++\right >, \qquad
{\bf H}_1 \left | --\right >=w_2\Bigl (1+\frac{\eta}{x_2-x_1}\Bigr )
\left | --\right >,
$$
$$
{\bf H}_1 \left | +-\right >=w_1 \left | +-\right >+
\frac{\eta w_1}{x_1-x_2}\left | -+\right >,
$$
$$
{\bf H}_1 \left | -+\right >=w_2 \left | -+\right >+
\frac{\eta w_2}{x_1-x_2}\left | +-\right >.
$$
Here we use the usual notation for the basis vectors 
in $\CC^2 \otimes \CC^2$:
$$\left | ++\right >=\left (\begin{array}{c}1\\0\end{array}\right )
\otimes \left (\begin{array}{c}1\\0\end{array}\right ), \qquad
\left | +-\right >=\left (\begin{array}{c}1\\0\end{array}\right )
\otimes \left (\begin{array}{c}0\\1\end{array}\right ), \quad
\mbox{and so on.}
$$
The vectors $\left | ++\right >$ and $\left | --\right >$
are eigenvectors of ${\bf H}_1$. The rest part of the spectrum is found by 
diagonalizing the $2\! \times \! 2$ matrix
$\displaystyle{\left (\begin{array}{cc}
w_1 & \frac{\eta w_1}{x_1-x_2}\\
\frac{\eta w_2}{x_1-x_2} & w_2\end{array}\right )}$.
The two eigenvalues are
$
\frac{1}{2}\left (w_1+w_2 \pm \sqrt{R} \right )$, where
$$R=(w_1-w_2)^2 +\frac{4\eta ^2 w_1 w_2}{(x_1-x_2)^2}.
$$
The final result for the joint spectrum of the operators 
${\bf H}_i$ is as follows:
\beq\label{HH}
(H_1, H_2)=\left \{
\begin{array}{l}
\displaystyle{ \left ( w_1+\frac{\eta w_1}{x_1-x_2}, \,\,
w_1-\frac{\eta w_1}{x_1-x_2} \right )}\,, \,\, \,\, \qquad M=0,
\\ \\
\displaystyle{\left ( \frac{w_1+w_2 +\sqrt{R}}{2}, \,\,
\frac{w_1+w_2 -\sqrt{R}}{2}\right )}\,, \! \! \! \! \qquad M=1,
\\ \\
\displaystyle{\left ( \frac{w_1+w_2 -\sqrt{R}}{2}, \,\,
\frac{w_1+w_2 +\sqrt{R}}{2}\right )}\,, \! \! \!\! \qquad M=1,
\\ \\
\displaystyle{ \left ( w_2+\frac{\eta w_2}{x_1-x_2}, \,\,\,
w_2-\frac{\eta w_2}{x_1-x_2} \right )}\,, \,\, \,\, \qquad M=2.
\end{array}\right. 
\eeq
Note that in the case of the periodic boundary condition 
$w_1=w_2=1$ the eigenvalue $H_1=1+\frac{\eta}{x_1-x_2}$ becomes
3-fold degenerate as it should be due to the $GL(2)$-invariance 
of the $R$-matrix.

Now consider the Lax matrix of the 2-particle RS model, where 
we substitute $\dot x_i =-H_i$:
$$
{\sf Y}=\left ( \begin{array}{cc}
H_1 & \displaystyle{\frac{\eta H_1}{x_2 \! - \! x_1\!  + \! \eta}}
\\ & \\
\displaystyle{\frac{\eta H_2}{x_1 \! - \! x_2\!  + \! \eta}} & H_2
\end{array} 
\right )
$$
The characteristic equation $\det ({\sf Y} -\lambda {\sf I})=0$ reads
$\displaystyle{
\lambda^2 -(H_1+H_2)\lambda +\frac{x_{12}^2 H_1H_2}{x_{12}^2
\! -\! \eta^2}=0}
$, where
$x_{12}\equiv x_1-x_2$ and the two eigenvalues are
$$
\frac{1}{2}\left ( H_1+H_2 \pm \sqrt{(H_1+H_2)^2 -
\frac{4x_{12}^2\, H_1H_2}{x_{12}^2
\! -\! \eta^2}}\,\, \right ).
$$
In the subspace with $M=0$ the eigenvalue of ${\bf H}_1+{\bf H}_2$ is $2w_1$
and the Lax matrix has the double eigenvalue $w_1$. This implies that 
the expression under the square root vanishes, i.e., we arrive at the system
$$
\left \{\begin{array}{l}
H_1\! + \! H_2=2w_1
\\
\displaystyle{H_1H_2=w_1^2 \Bigl (1-\frac{\eta^2}{x_{12}^2}\Bigr )}
\end{array}
\right.
$$
which is a particular case $N=2$ of the general system (\ref{QC7}).
There are two solutions: 
$$(H_1, H_2)=\left ( w_1 \pm \frac{\eta w_1}{x_1 \! -\!  x_2}, \,\,
w_1 \mp \frac{\eta w_1}{x_1 \! -\!  x_2}\right )\,, \qquad M=0.
$$
The choice of the upper sign corresponds to the first line in (\ref{HH}).
The meaning of the other solution is to be clarified.
In a similar way, for $M=2$ we obtain two solutions
$$(H_1, H_2)=\left ( w_2 \pm \frac{\eta w_2}{x_1 \! -\!  x_2}, \,\,
w_2 \mp \frac{\eta w_2}{x_1 \! -\!  x_2}\right )\,, \qquad M=2,
$$
of which the one with the upper sign corresponds to the last line in
(\ref{HH}). Finally, at $M=1$ we have the system
$$
\left \{\begin{array}{l}
H_1\! + \! H_2=w_1 +w_2
\\
\displaystyle{H_1H_2=w_1w_2 \Bigl (1-\frac{\eta^2}{x_{12}^2}\Bigr )}.
\end{array}
\right.
$$
There are two solutions which coincide with the second and the third
lines in (\ref{HH}).

\section{The limit to the 
quantum Gaudin model and the classical Calogero-Moser system}

In the limit $\eta \to 0$ the QC duality discussed above becomes
a correspondence (\ref{three}) between the quantum Gaudin model 
and the classical Calogero-Moser system with inversely 
quadratic pair potential. Some details 
are given below.

The rational $GL(2)$ Gaudin model \cite{Gaudin} is the $\eta \to 0$ limit 
of the inhomogeneous spin chain with the transfer matrix
${\bf T}(x; e^{\eta h} , \eta , \{x_j \})$.
The expansion as $\eta \to 0$ gives:
$$
{\bf T}(x;  e^{\eta h}, \eta , \{x_j \})=
2{\bf I}+\eta \left ( \mbox{tr} \, h +\sum_{i=1}^N \frac{1}{x-x_i}\right )
{\bf I}
+\eta^2 \left ( \frac{1}{2} \, \mbox{tr} \, h^2 \, {\bf I} +\sum_{i=1}^N
\frac{{\bf H}_{i}^G}{x-x_i} \right ) +O(\eta^3),
$$
where $h=\left (\begin{array}{cc}\omega _1 & 0\\ 0& \omega_2 \end{array}\right )$
is the Gaudin analogue of the twist matrix, and 
\beq\label{G1}
\begin{array}{c}\displaystyle{
{\bf H}_{i}^G=\lim_{\eta \to 0}\, \frac{{\bf H}_{i}(\eta ,
e^{\eta h}, \{x_j\})-{\bf I}}{\eta}=
{\bf h}^{(i)}+\sum_{j\neq i} \frac{{\bf P}_{ij}}{x_i-x_j}}
\\ \\
\displaystyle{=\,\, \sum_{j\neq i}\frac{{\bf I}}{x_i-x_j} \, +\, {\bf h}^{(i)}+
\, 2\sum_{j\neq i}
\frac{\vec {\bf s}^{(i)}\vec {\bf s}^{(j)}}{x_i-x_j}}
\end{array}
\eeq
are the Hamiltonians of the $GL(2)$-invariant Gaudin model.
Here
$
{\bf h}^{(i)} =\frac{\omega_1+\omega_2}{2}\, {\bf I} +
(\omega_1-\omega_2){\bf s}_z^{(i)}
$
is the twist matrix acting in the space $V_i \cong \CC^2$ at the $i$th site.
In the context of the Gaudin model, the parameters $x_i$ 
(in general, complex numbers) are often called marked points
of the Riemann sphere. Since the first two terms in the $\eta \to 0$
expansion of the ${\bf T}(x;  e^{\eta h}, \eta , \{x_j \})$ are
proportional to the identity operator and thus commute with everything,
commutativity of the transfer matrices implies commutativity 
of the Gaudin Hamiltonians: $[{\bf H}_{i}^G, \, {\bf H}_{j}^G]=0$.
The Gaudin spectral problem consists in the simultaneous diagonalization 
of these operators and the operator ${\bf M}$ which has the same form as 
above: ${\bf H}_{i}^G\Psi =H_i^G \Psi$, ${\bf M}\Psi = M\Psi$.
The Bethe ansatz solution is the $\eta \to 0$ limit of 
(\ref{J6}), (\ref{Bethe3}):
\beq\label{G2}
H_j^G=\omega_1 +\sum_{k\neq j}\frac{1}{x_j-x_k}+
\sum_{\alpha =1}^M \frac{1}{u_\alpha -x_j},
\eeq
where the Bethe roots $u_{\alpha}$ satisfy the system of equations
\beq\label{G3}
\omega_1 -\omega_2 +\sum_{k=1}^N \frac{1}{u_{\alpha}-x_k} =
2\sum_{\beta =1, \neq \alpha}^M  \frac{1}{u_{\alpha}-u_{\beta}}.
\eeq

An alternative solution is achieved via the QC duality 
with the classical CM model with the Hamiltonian
$\displaystyle{
{\cal H}^{\rm CM}=\frac{1}{2}\sum_{i=1}^N p_i^2 -\sum_{i<j}
\frac{1}{(x_i-x_j)^2}}$.
The equations of motion are
\beq\label{CM2}
\ddot x_i=-\sum_{k\neq i}\frac{2}{(x_i-x_k)^3}, \qquad i=1, \ldots , N.
\eeq
The CM model is known to be integrable, with the higher integrals
of motion in involution being given by ${\cal H}^{\rm CM}_k=\frac{1}{k}
\mbox{tr}\, ({\sf Y}^{\rm CM})^k$ (${\cal H}_1^{\rm CM}$
being the total momentum ${\cal P}^{\rm CM}=
\sum_j p_j$
and ${\cal H}_2^{\rm CM} ={\cal H}^{\rm CM}$), where 
\beq\label{Lax2}
{\sf Y}^{\rm CM}(\{x_i\}; \{\dot x_i \})=\left ( \begin{array}{ccccc}
\displaystyle{-\dot x_1} & \displaystyle{\frac{1}{x_2-x_1}} 
&\displaystyle{\frac{1}{x_3-x_1}} &
\ldots & \displaystyle{\frac{1}{x_N-x_1}}
\\ &&&& \\
 \displaystyle{\frac{1}{x_1-x_2}} & 
 \displaystyle{-\dot x_2}& 
 \displaystyle{\frac{1}{x_3-x_2}} &
 \ldots & \displaystyle{\frac{1}{x_N-x_2}}
 \\ &&&& \\ \vdots & \vdots & \vdots & \ddots & \vdots
 \\ &&&& \\
 \displaystyle{\frac{1}{x_1-x_N}} & 
 \displaystyle{\frac{1}{x_2-x_N}}&
 \displaystyle{\frac{1}{x_3-x_N}} & \ldots & 
 \displaystyle{-\dot x_N}
\end{array}\right )
\eeq
is the Lax matrix of the model. Its matrix elements are
$\displaystyle{{\sf Y}^{\rm CM}_{ij} 
=-\dot x_i \delta_{ij}-\frac{1-\delta_{ij}}{x_i-x_j}}$.

Note that the CM model can be treated as a $\eta \to 0$ limit of the RS model
meaning that
$$
{\sf Y}^{\rm RS}={\sf I}+\eta {\sf Y}^{\rm CM}+O(\eta ^2)\,, \qquad
{\cal H}_1^{\rm RS}=\frac{N}{\eta}+{\cal P}^{\rm CM} +\eta \tilde {\cal H}^{\rm CM}+
O(\eta ^2),
$$
where $\displaystyle{\tilde {\cal H}^{\rm CM}=\frac{1}{2}\sum_{i}\Bigl (
p_i +\sum_{k\neq i}\frac{1}{x_i-x_k}\Bigr )^2-\sum_{i<j}^N
\frac{1}{(x_i-x_j)^2}}$ differs from the ${\cal H}^{\rm CM}$ by a simple canonical
transformation and leads to the same equations of motion.

The rules of the QC duality in this case are as follows \cite{ALTZ13,GZZ14}.
Consider the Lax matrix (\ref{Lax2}) of the 
$N$-particle CM model, where the $x_i$'s
are identified 
with the $N$ marked points of the Gaudin model.
Let us also substitute 
$\dot x_i =-H^G_i$:
\beq\label{CM3}
{\sf Y}^{\rm CM}(\{x_i\}; \{-H_i\})
=\left ( \begin{array}{ccccc}
\displaystyle{H^G_1} & \displaystyle{\frac{1}{x_2\! -\!x_1}} 
&\displaystyle{\frac{1}{x_3\! -\! x_1}} &
\ldots & \displaystyle{\frac{1}{x_N\! -\! x_1}}
\\ &&&& \\
 \displaystyle{\frac{1}{x_1\! -\! x_2}} & 
 \displaystyle{H^G_2}& 
 \displaystyle{\frac{1}{x_3\! -\! x_2}} &
 \ldots & \displaystyle{\frac{1}{x_N\! -\! x_2}}
 \\ &&&& \\ \vdots & \vdots & \vdots & \ddots & \vdots
 \\ &&&& \\
 \displaystyle{\frac{1}{x_1\! -\! x_N}} & 
 \displaystyle{\frac{1}{x_2\! -\! x_N}}&
 \displaystyle{\frac{1}{x_3\! -\! x_N}} & \ldots & 
 \displaystyle{H^G_N}
\end{array}\right ).
\eeq
The claim is that if the $H_i^G$'s are eigenvalues of the Gaudin Hamiltonians
in the invariant subspace ${\cal V}(M)$, then 
the first $N-M$ eigenvalues of this matrix coincide with
eigenvalues of the twist matrix $\omega_1$ 
while the rest $M$ eigenvalues 
coincide with $\omega_2$:
\beq\label{CM4}
\mbox{Spec}\, ({\sf Y}^{\rm CM})=\Bigl (
\underbrace{\omega_1, \ldots , \omega_1}_{N-M}, \,
\underbrace{\omega_2, \ldots , \omega_2}_{M} \Bigr ).
\eeq

As it follows from the results of \cite{Sawada-Kotera,Woj},
the characteristic polynomial of the matrix ${\sf Y}^{\rm CM}$ can be 
represented in the form
\beq\label{CM5}
\left.
\det_{N\times N} \left (\lambda {\sf I}-{\sf Y}^{\rm CM}\right )=
\exp \Bigl ( \sum_{i<j}^N \frac{\p_{y_i}\p_{y_j}}{(x_i\! -\! x_j )^2}\Bigr )
\prod_{k=1}^N (\lambda - y_k)\right |_{y_i=H_i^G}.
\eeq
Therefore, the spectrum consists of the values $(H_1, H_2, \ldots , H_N)$
such that the equality
\beq\label{CM6}
\left.
\exp \Bigl ( \sum_{i<j}^N \frac{\p_{y_i}\p_{y_j}}{(x_i\! -\! x_j )^2}\Bigr )
\prod_{k=1}^N (\lambda - y_k)\right |_{y_i=H_i^G}\!\!\!
=\,\, (\lambda -w_1)^{N-M}(\lambda -w_2)^M
\eeq
is satisfied identically in $\lambda$. As in the case 
of the XXX model, this is equivalent to $N$ 
algebraic equations for $N$ quantities $H^G_i$.

\section{Concluding remarks}

The QC duality can be more or less straightforwardly extended to 
quantum inhomogeneous spin chains 
associated with $GL(n)$-invariant $R$-matrices. 
These models are solved by the nested Bethe ansatz (see \cite{KR83}).
On the classical side, the correspondence 
is with the same rational RS model, with eigenvalues of the 
Lax matrix being chosen (with some multiplicities) from the elements of 
the $n\! \times \! n$ diagonal twist 
matrix. The corresponding results 
can be found in \cite{AKLTZ11,Zsigma,GZZ14}.
In the present paper, 
we have restricted ourselves by the $GL(2)$ case only because of 
the notational simplicity. 

An interesting possible generalization is the $q$-deformation of the 
QC duality which implies the anisotropic spin chains  
with trigonometric $R$-matrices (associated with $U_q(gl_n)$)
on the quantum side. As is shown in \cite{Z12}, the classical side
in this case is represented by the trigonometric RS model.
However, some interesting details, including an accurate
limit to the trigonometric Gaudin model, are still to be elaborated.

Among future perspectives we mention
an extension to the supersymmetric $GL(n|m)$-invariant 
spin chains and to the spin chains
with elliptic $R$-matrices. The latter case seems to be especially non-trivial
because integrable magnets constructed with the help of elliptic
$R$-matrices do not allow twisted boundary conditions with continuous 
parameters. That is why it is not clear how to fix values of the 
classical integrals of motion in the elliptic RS model which would be
the most natural candidate for the classical part of the QC duality.
Another difficulty is that the Lax matrix for the elliptic RS model 
contains a spectral parameter. The role of this parameter in the 
context of the quantum-classical correspondence is not clear at the 
moment.

\section*{Acknowledgements}

Discussions with A.Alexandrov,
A.Gorsky, V.Kazakov, S.Khoroshkin, I.Krichever, S.Leu\-rent,
M.Ol\-sha\-nets\-ky, A.Orlov, T.Ta\-ke\-be, Z.Tsuboi, and A.Zo\-tov 
are gratefully acknowledged. 
Some of these results were reported 
at the International School and Workshop ``Nonlinear Mathematical Physics and Natural
Hazards'' (November 28 - December 2 2013, Sofia, Bulgaria). 
The author thanks the organizers and especially professors
B.Aneva and V.Gerdzhikov for the invitation and support.
This work was supported in part
by RFBR grant 12-01-00525, by joint RFBR grants 
12-02-91052-CNRS, 14-01-90405-Ukr and
grant NSh-1500.2014.2 for support of
leading scientific schools.

}


\begin{thebibliography}{99}


\bibitem{AKLTZ11} A. Alexandrov, V. Kazakov, S. Leurent,
Z. Tsuboi and A. Zabrodin, {\it Classical tau-function for quantum
spin chains}, JHEP {\bf 1309} (2013) 064 [arXiv:1112.3310].

\bibitem{ALTZ13} A. Alexandrov, S. Leurent,
Z. Tsuboi and A. Zabrodin, {\it The master $T$-operator
for the Gaudin model and the KP hierarchy}, Nucl. Phys.
{\bf B883} (2014) 173-223 [arXiv:1306.1111].

\bibitem{Zsigma}  A. Zabrodin, {\it The master T-operator for 
inhomogeneous XXX spin chain and mKP hierarchy} SIGMA {\bf 10} (2014)
006 (18 pages), [arXiv:1310.6988].

\bibitem{GZZ14} A. Gorsky, A. Zabrodin and A. Zotov, {\it Spectrum
of quantum transfer matrices via classical many-body systems}, JHEP
{\bf 01} (2014) 070, [arXiv:1310.6958]. 

\bibitem{RS} S.N.M. Ruijsenaars and H. Schneider, {\it A new class
of integrable systems and its relation to solitons}, 
Ann. Phys. {\bf 170} (1986) 370-405;\\
S.N.M. Ruijsenaars, {\it Complete integrability of relativistic
Calogero-Moser systems and elliptic function identities},
Commun. Math. Phys. {\bf 110} (1987) 191-213.


\bibitem{Calogero}
F. Calogero, {\it Solution of the one-dimensional $N$-body problems
with quadratic and/or inversely quadratic pair potentials},
J. Math. Phys. {\bf 12} (1971) 419-436;\\
J. Moser, {\it Three integrable hamiltonian systems connected with
isospectrum deformations}, Adv. Math. {\bf 16} (1976) 354-370.




\bibitem{OP} M. Olshanetsky and A. Perelomov, {\it Classical
integrable finite dimensional systems related to Lie algebras},
Phys. Reps. {\bf 71} (1981) 313-400.

\bibitem{Gaudin} M.~Gaudin, {\it Diagonalisation d'une classe 
d'hamiltoniens de spin}, J. de Phys. {\bf 37} (1976), no. 10 1087-1098.

\bibitem{GK} A. Givental and B.-S. Kim, {\it Quantum cohomology
of flag manifolds and Toda lattices}, Commun. Math. Phys.
{\bf 168} (1995) 609-641 [arXiv:hep-th/9312096].

\bibitem{MTV09} E. Mukhin, V. Tarasov and A. Varchenko,
{\it Gaudin Hamiltonians generate the Bethe algebra
of a tensor power of vector representation of $gl_N$},
St. Petersburg Math. J. {\bf 22} (2011) 463-472
[arXiv:0904.2131];\\
E. Mukhin, V. Tarasov and A. Varchenko, 
{\it Bethe algebra of Gaudin model, Calogero-Moser space and Cherednik algebra},
Int. Math. Res. Not. {\bf 2014} (2014) Issue 5 1174-1204
[arXiv:0906.5185].
   
    
\bibitem{MTV12}
E. Mukhin, V. Tarasov and A. Varchenko,
{\it KZ characteristic variety as the zero set of
classical Calogero-Moser Hamiltonians},  SIGMA {\bf 8} (2012)
072 (11 pages)
[arXiv:1201.3990];\\
E. Mukhin, V. Tarasov and A. Varchenko, {\it Bethe subalgebras
of the group algebra of the symmetric group}, [arXiv:1004.4248];\\
E. Mukhin, V. Tarasov and A. Varchenko, {\it Spaces of 
quasi-exponentials and representations of the Yangian
$Y(gl_N)$}, [arXiv:1303.1578].



\bibitem{Z12} A. Zabrodin, {\it The master $T$-operator for vertex models with trigonometric $R$-matrices as classical tau-function}, Teor. Mat. Fys.
{\bf 171:1} (2013) 59-76 (Theor. Math. Phys. {\bf 174} (2013) 52-67)
[arXiv:1205.4152]

\bibitem{Zjapan}
A. Zabrodin, {\it Hirota equation and Bethe ansatz in integrable models},
Suuri-kagaku Journal (in Japanese),
Number 596 (2013) 7-12.



\bibitem{DJKM83} E. Date, M. Jimbo, M. Kashiwara and T. Miwa,
{\it Transformation groups for soliton equations},
in ``Nonlinear integrable systems -- classical and quantum'',
eds. M. Jimbo and T. Miwa, World Scientific, pp. 39-120 (1983);\\
M. Jimbo and T. Miwa, {\it Solitons and
infinite dimensional Lie algebras}, Publ. RIMS, Kyoto Univ.
{\bf 19} (1983) 943-1001.


\bibitem{NRS} N. Nekrasov, A. Rosly and S. Shatashvili,
``Darboux coordinates, Yang-Yang functional, and gauge theory'',
Nucl. Phys. Proc. Suppl. {\bf 216} (2011) 69-93 [arXiv:1103.3919].

\bibitem{GK13}
 D. Gaiotto and P. Koroteev,
    {\it On three dimensional quiver gauge theories and integrability},
     JHEP {\bf 05} (2013) 126 
   [arXiv:1304.0779].


\bibitem{KLWZ97}
I.\ Krichever, O.\ Lipan, P.\ Wiegmann and A.\ Zabrodin,
{\it Quantum Integrable Models and Discrete Classical Hirota Equations},
Commun. Math. Phys. {\bf 188} (1997) 267-304 [arXiv:hep-th/9604080];\\
A. Zabrodin, {\it
Discrete Hirota's equation in quantum integrable models},
Int. J. Mod. Phys. {\bf B11} (1997) 3125-3158;\\
A. Zabrodin, {\it Hirota equation and Bethe ansatz},
Teor. Mat. Fyz., {\bf 116} (1998) 54-100
(English translation:
 Theor. Math. Phys. {\bf 116}
(1998) 782-819.

\bibitem{KSZ08}
V.~Kazakov, A.~S.~Sorin and A.~Zabrodin,
{\it Supersymmetric Bethe ansatz and Baxter 
equations from discrete Hirota dynamics},
Nucl.\ Phys.\ B {\bf 790} (2008) 345-413
[arXiv:hep-th/0703147];\\
A. Zabrodin, {\it B\"acklund transformations for
difference Hirota equation and
supersymmetric Bethe ansatz},
Teor. Mat. Fyz. {\bf 155} (2008) 74-93 (English
translation: Theor. Math. Phys. {\bf 155} (2008) 567-584)
[arXiv:0705.4006].








\bibitem{Bethe} H. Bethe, {\it Zur Theorie der Metalle. I.
Eigenwerte und Eigenfunktionen der linearen Atomkette}, 
Zeitschr. fur Physik {\bf 71} (1931) 205-226.




\bibitem{QISM1} L. Faddeev, E. Sklyanin and L. Takhtajan,
{\it The quantum inverse problem method. I}, Theor. Math. Phys.
{\bf 40} (1980) 688;\\
V.E. Korepin, N.M. Bogoliubov and A.G. Izergin,
{\it Quantum inverse scattering method and correlation functions},
Cambridge Monographs on Mathematical Physics, Cambridge University
Press, Cambridge U.K., 1997.

\bibitem{FTLOMI} L. Faddeev and L. Takhtajan, 
{\it The spectrum and scattering of excitations in the 
one-dimensional isotropic Heisenberg model}, Zap. Nauch. Semin.
LOMI {\bf 109} (1981) 134-178.




\bibitem{HKW92} K. Hikami, P. Kulish and M. Wadati, {\it 
Construction of integrable spin systems with long-range 
interaction}, J. Phys. Soc. Japan {\bf 61} (1992) 3071-3076.







  

\bibitem{Macdonald} I. Macdonald, {\it Symmetric functions and
Hall polynomials}, 2nd ed., Oxford University Press, 1995.

\bibitem{Sawada-Kotera} K. Sawada and T. Kotera, 
{\it Integrability and a solution for the one-dimensional $N$-particle
system with inversely quadratic pair potential}, J. Phys. Soc.
Japan, {\bf 39} (1975) 1614-1618.

\bibitem{Woj} S. Wojciechowski, {\it New completely integrable 
Hamiltonian systems of $N$ particles on the real line}, Phys. Lett. 
{\bf 59A} (1976) 84-86.

\bibitem{KR83} P. Kulish and N. Reshetikhin, {\it 
Diagonalization of $gl_N$ invariant transfer matrices
and quantum $N$-wave system (Lee model)}, J. Phys. A {\bf 16} (1983)
L591-L596.


\end{thebibliography}
\end{document}